# High-resolution chemical and structural characterization of the native oxide scale on a Mg-based alloy

Deborah Neuß[1]*, Ingrid E. McCarroll[2], Siyuan Zhang[2], Eric Woods[2], Wassilios J. Delis[3], Leandro Tanure[4], Hauke Springer[4], Stefanie Sandlöbes[3], Jing Yang[2], Mira Todorova[2], Daniela Zander[5], Christina Scheu[2], Jochen M. Schneider[1], Marcus Hans[1]

[1]Materials Chemistry, RWTH Aachen University, Kopernikusstraße 10, 52074 Aachen, Germany

[2]Max-Planck-Institut für Eisenforschung GmbH, Max-Planck-Straße 1, 40237 Düsseldorf, Germany

[3]Institute for Physical Metallurgy and Materials Physics, RWTH Aachen University, Kopernikusstraße 14, 52074 Aachen, Germany

[4]Metallic Composite Materials, RWTH Aachen University, Intzestraße 10, 52072 Aachen, Germany Aachen, Germany

[5]Chair of Corrosion and Corrosion Protection, RWTH Aachen University, Intzestraße 5, 52072 Aachen, Germany

*Corresponding author; E-Mail: neuss@mch.rwth-aachen.de

**Abstract**

The structure and composition of the native oxide forming on the basal plane (0001) of the magnesium-based alloy Mg-2Al-0.1Ca is investigated by combining scanning transmission electron microscopy (STEM) and atom probe tomography (APT). Site-specific atom probe specimens are extracted conventionally with a gallium focused ion beam (FIB) as well as with a Xe plasma FIB in a cryogenic setup. After a cleaning step the specimens are exposed to air under defined conditions to form the native oxide scale, which is subsequently analyzed by STEM and APT. STEM measurements demonstrate the growth of a (111) MgO oxide layer with 3-4 nm thickness on the basal (0001) plane of the atom probe specimen. While longer atmospheric exposure times increase the measured hydrogen content in the specimen's apex region, the comparison between conventional and cryogenic preparation reveals, that the hydrogen uptake in magnesium is independent of the preparation strategy and appears to be governed by residual gas in the analysis chamber. APT data further reveals the formation of an aluminum-rich region between bulk magnesium and the native oxide. The aluminum enrichment of up to ~20 at.% at the interface is consistent with an inward growth of the oxide scale.

# 1. Introduction

Magnesium is an attractive metal in lightweight materials design, such as automotive or aerospace applications, due to its high strength-to-mass ratio [1-7]. Besides structural properties, the chemical stability of magnesium-based alloys is of paramount importance in corrosive application environments. It is well known that non-protective magnesium oxide scales are formed at high temperatures in oxygen-containing atmospheres [8, 9]. Therefore, the oxidation properties and especially the relevant mechanisms during oxidation need to be understood for the design of novel magnesium-based alloys. According to Fournier *et al.* [10], the oxide scale on magnesium at room temperature reaches 1.5 nm for 15 min oxidation in pure oxygen atmosphere at 20 kPa, while the oxide scale thickness increases to 2.6 nm at 300 °C under otherwise identical conditions. Comparing oxidation rates of magnesium and magnesium alloys, it is indicated that 400 °C is the critical onset temperature for the growth of a non-protective oxide layer [11]. In aqueous solutions, magnesium is dissolved by an electrochemical reaction with water that produces $Mg(OH)_2$ and gaseous $H_2$ [12].

It has been reported that the corrosion resistance of magnesium is also affected by the texture. According to Liu *et al.* [13], the (0001) basal plane is the most corrosion resistant, while the prismatic planes $(01\bar{1}0)$ and $(11\bar{2}0)$ exhibit larger surface energies and are hence less corrosion resistant. A magnesium sample was corroded for 15 h in aqueous 0.1 M HCl at room temperature, resulting in an oxide scale of 0.87 mm and the maximum difference in corrosion depth between two grains was about 10 % of the total corrosion depth [13].

Furthermore, the chemical composition of magnesium-based alloys has a significant influence on the oxidation behavior. It has been shown in several studies that the addition of small amounts of aluminum and calcium improves the oxidation resistance of magnesium-based alloys considerably [14-17]. Drynda *et al.* [15] studied the corrosion in experiments with 2.5% NaCl and it was demonstrated that additions of up to 0.8 wt.% calcium decrease the corrosion rate from ~790 µA cm$^{-2}$ to ~490 µA cm$^{-2}$. It was suggested that the higher affinity of calcium to oxidize leads to the formation of an outer oxide shell which acts as a protective layer during further exposure [8, 18]. While the presence of calcium in the outermost oxide shell was shown by Auger electron spectroscopy [19, 20], the thickness of the calcium oxide layer has not been quantified. Aluminum has also been shown to have a beneficial effect on the corrosion properties of magnesium alloys. The work of Mathieu *et al.* [17] revealed that the corrosion potential of magnesium-based solid solutions is dependent on the aluminum content: While magnesium has a corrosion potential of -1.55 V, there is a linear rise up to -1.40 V with the addition of 9 at.% aluminum [17]. According to Hermann *et al.* [21], the enhanced chemical stability can be attributed to the formation of a protective aluminum-enriched interdiffusion zone between surface and bulk. The existence of this zone was suggested based on an increase in pitting potential with increasing aluminum content, as well as an increase in selectivity and the virtual cessation of macroscopic aluminum dissolution [21]. Furthermore, Jeurgens *et al.* [22] suggested the formation of an aluminum-enriched zone in aluminum concentrations of 2.6, 5.8 and 7.3 at.% due to argon sputter cleaning to remove adventitious carbon during an X-ray photoelectron spectroscopy measurement. It was assumed that the segregation was induced by bombardment-enhanced Gibbsian segregation of magnesium from the inner alloy towards the bombarded surface region [22].

Based on the discussion above, most studies have considered either oxidation at high temperatures or in immersion testing, while less attention has been paid to the native oxide which is formed at the alloy surface due to air exposure under ambient conditions. In order to understand the oxidation behavior and the underlying mechanisms in magnesium-based alloys, it is vital to characterize the initial structure and composition of the native oxide prior to further corrosion. As these oxide scales are nanometer-sized, advanced microscopy techniques at (near-)atomic resolution are necessary to unravel the structure and composition of the native oxide.

Atom probe tomography (APT) combines mass spectrometry and projection microscopy for spatially-resolved chemical composition analysis of nanometer-sized features [23, 24]. Individual atoms are removed from needle-shaped, cryogenically cooled specimens based on field evaporation and post-ionized in the presence of an electric field [25, 26]. The surface electric field is created by applying a so-called standing voltage to the specimen and field evaporation is controlled by either voltage or laser pulsing to overcome the evaporation threshold. The formed elemental or molecular ions are accelerated towards a position-sensitive detector, while the time-of-flight is measured for determining the chemical identity [23, 27, 28]. Thereby, three-dimensional images of chemical composition distributions are created with spatial resolution at the nanometer scale.

Even though the opportunities provided by APT for sample analysis on a small scale are widely recognized, instrumental challenges need to be considered. In the literature, the size of the smallest microstructural object that can be analyzed by APT is debated and the limiting factor for the accurate and precise characterization of individual objects is the effective spatial resolution [29]. The spatial resolution is mainly affected by ion trajectory aberrations in the early stages of flight [29]. Multiple detection events can occur, if more than one atom is evaporated from the probed area during a single pulse and those species arrive simultaneously at the position-sensitive detector [30]. A non-uniform distribution of the electric field at the surface area of the specimen can cause this effect [30]. Furthermore, a single evaporation event might trigger the evaporation of surrounding atoms due to the modified surface roughness [30]. These challenges can be partly overcome by correlating the composition at the nanometer scale from APT with the nanostructure information from scanning transmission electron microscopy (STEM).

In the present work we have characterized the structure and composition of the native oxide formed on a Mg-2Al-0.1Ca alloy combining STEM and APT at an identical specimen. Needle-shaped specimens were prepared from a grain with (0001) surface and it is demonstrated that MgO with 3 to 4 nm thickness and (111) orientation is formed on the (0001) Mg surface upon exposure to ambient atmosphere. An aluminum-rich region forms between bulk and native oxide upon exposure to ambient air as revealed by both laser-assisted and voltage field evaporation. The aluminum enrichment of up to ~20 at.% at the interface between bulk and native oxide is consistent with the inward growth of the oxide scale. Effects of specimen preparation route, detection efficiency as well as voltage and thermal pulsing on the quantification by APT are discussed. It is demonstrated that the presented approach, combining STEM and APT enables identification of the structure and composition of the native oxide scale. Our findings provide a baseline for studying corrosion of magnesium-based alloys during high temperature oxidation and immersion testing.

## 2. Methods

The Mg-2Al-0.1Ca sample was produced by induction melting under an argon atmosphere of 10 bar to limit both evaporation and oxidation. Charges of ~300 g of raw materials were inductively molten in a steel crucible and cast into rectangular copper molds (internal cross-section of 30 × 150 mm$^2$). After casting the material was heated to 450°C for 30 minutes and hot-rolled aiming for 50% of thickness reduction (10% of reduction per pass, reheating for 10 min between passes, and final reheating for 15 min followed by water quenching). Chemical homogeneity was enhanced by an additional annealing heat treatment at 500°C for 24 hours, and once again, water quenched afterwards. Wet chemical analysis determined the actual composition as Mg-2.11Al-0.11Ca (at.%), with Cu and Ni impurities < 0.002 at.%.

For electron backscatter diffraction (EBSD) the sample was mechanically ground and then polished with 0.25 µm diamond suspension. The preparation was finished with an electropolishing step with the AC2 electrolyte of Struers (30 V, 30 seconds at -20 °C). EBSD was performed with a FEI Helios Nanolab 600i at 70° tilt angle, 20 kV acceleration voltage, and 5.5 nA current.

Recently, STEM and APT have been correlatively applied to investigate the oxide scale of an Fe-13Cr (at.%) alloy [31] and similar protocols have been employed in the present work. All atom probe specimens from the Mg-2Al-0.1Ca alloy were extracted through site-specific lift-out from a grain with (0001) surface using focused ion beam (FIB) techniques in a FEI Helios Nanolab 660 dual-beam microscope with gallium ions. As it is well known that gallium is incorporated into magnesium-containing specimens during FIB preparation [32, 33], annular milling was finished with a 5 kV cleaning step [34]. Furthermore, the gallium-containing top part of the specimen was removed using APT in order to generate a damage-free, pristine surface according to the approach presented by Sasidhar *et al.* [31]. Laser-assisted field evaporation was carried out within a CAMECA local electrode atom probe (LEAP) 4000X HR until the average gallium concentration was < 1 at.%. The pristine surfaces were introduced to ambient air at a temperature of 20 °C and relative humidity of 40 % for 12 hours, followed by STEM and APT characterization.

3D spatially-resolved analysis of the chemical composition was carried out by APT using different microscopes, i.e. LEAP 4000X HR for laser-assisted thermal pulsing and a LEAP 5000 XS for voltage pulsing. While the LEAP 4000X HR has a reflectron for enhanced mass resolution, the detection efficiency is limited to 36%. In contrast, the LEAP 5000 XS has a detection efficiency of 86% due to the straight flight path. For simplicity, the different type of APT measurements will be referred to as laser mode (LEAP 4000) and voltage mode (LEAP 5000) in the following. Measurements in laser mode were carried out with 60 pJ pulse energy, 125 kHz pulse frequency, and an average detection rate of 0.5 %, while in voltage mode 20% pulse fraction, 125 kHz pulse frequency, and an average detection rate of 1% were employed at 30 K base temperature.

In addition, a magnesium reference sample was also investigated after exposure to ambient air for 11 days to compare the composition of the native oxide scale to the Mg-2Al-0.1Ca alloy. Moreover, for investigation of preparation-induced hydrogen uptake, a preparation route at cryogenic temperatures was chosen using a Helios dual-beam plasma FIB (PFIB), employing xenon ions for sharpening of the specimens. These specimens were transferred using a cryo vacuum transfer system including a FerroVac vacuum suitcase, that can be attached to the PFIB as well as to the atom probe. Thereby, hydrogen incorporation during specimen preparation as well as transport to the atom probe microscope was minimized. This specimen was measured with voltage pulsing using 15% pulse fraction, 125 kHz pulse frequency, and average detection rate of 3%, while the base temperature was 30 K.

Specimens for STEM analysis were prepared on titanium grids. These grids were sectioned in halves using a razor blade and the 50 µm wide posts were pre-thinned using the gallium FIB to cones with approximately 2 µm diameter at the top end for attachment of atom probe specimens following the protocol introduced by Herbig *et al*. [35]. STEM measurements were conducted at 300 kV with a Thermo Fisher Scientific Titan Themis STEM 80-300 equipped with a Cs probe corrector. The convergence angle and size of the electron beam were 23.8 mrad and 0.1 nm, respectively. Images were taken with the high angle annular dark field (HAADF) and annular bright field (ABF) detectors with collection angles of 73-200 and 8-16 mrad, respectively. Elemental mapping was done using energy dispersive X-ray spectroscopy (EDX) collected using a Super X-detector. Multivariate statistical analysis was performed on the EDX spectrum imaging dataset for noise reduction [36] before quantification was performed on each spectral component using the Cliff-Lorimer method.

## 3. Results and Discussion

3.1. Site-specific lift-out and specimen cleaning

The microstructure of the Mg-2Al-0.1Ca alloy is shown in Figure 1 together with an inverse pole figure, obtained from EBSD mapping. Equiaxed grains with diameters between 10 and 200 µm are visible and a ~100 µm wide grain with (0001) surface (2° misorientation) was chosen for atom probe specimen preparation. All of the investigated Mg-2Al0.1Ca specimens reported in this article were extracted from this grain.

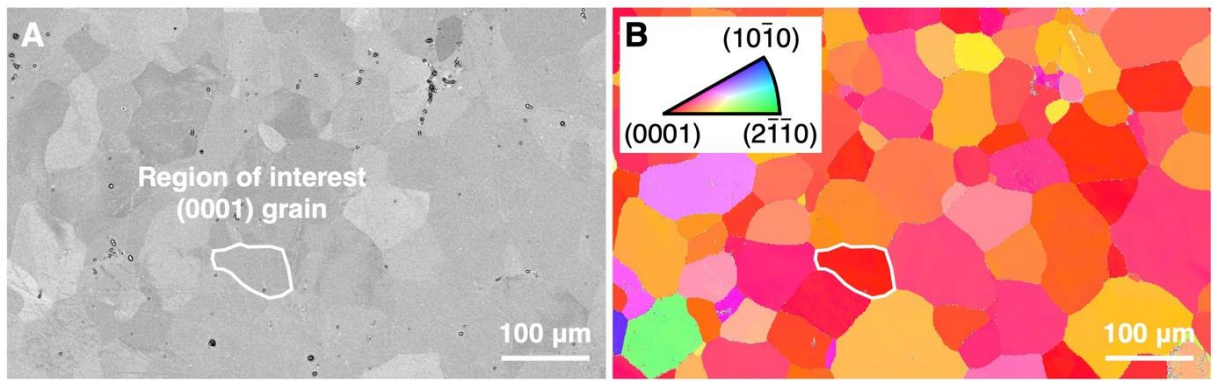

**Figure 1.** (a) Backscatter electron image where the region of interest is marked. (b) Inverse pole figure from EBSD mapping. A grain with (0001) surface was chosen based on the corresponding inverse pole figure from EBSD mapping.

A representative reconstruction of an atom probe specimen during the cleaning step is shown in Figure 2. From the exposure to the ion beam in the gallium FIB, penetration of gallium ions to a certain extent is inevitable. Hence, a cleaning step is necessary to remove as much gallium as possible, generating a clean surface and minimizing the influence of incorporated gallium on the native oxide formation. While the first 150 nm of the specimen's apex region contain up to 7 at.% gallium, the center of the specimen shows a gallium content below 1 at.% for depths > 150 nm. After evaporation of 350 nm the incorporated gallium along the shank of the specimen is also removed. Furthermore, it is obvious that this cleaning step generates a larger top surface area on the specimen's apex which facilitates the subsequent structural and compositional characterization of the native oxide scale.

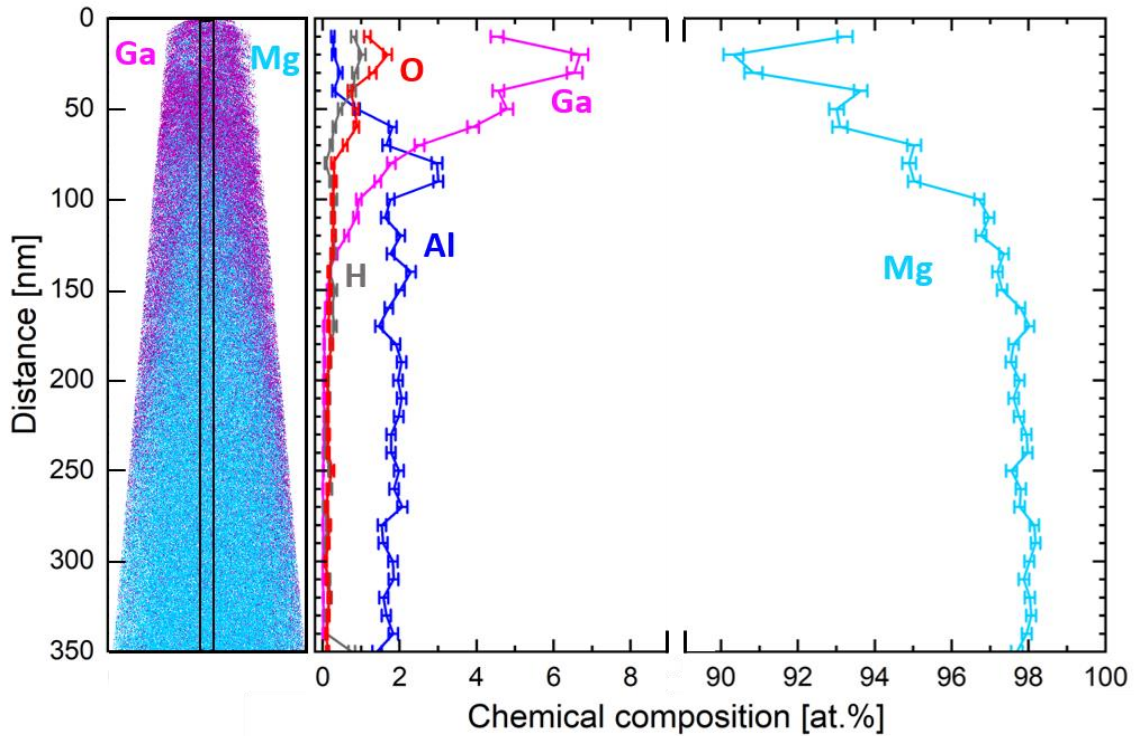

**Figure 2.** Reconstructed APT specimen of Mg-2Al-0.1Ca during the gallium cleaning step with an indication of specimen length and corresponding chemical composition along the drawn-in 350x10x10 nm cylinder.

Besides preparation-induced gallium incorporation, hydrogen contamination is an issue and challenges the quantification thereof within atom probe specimens. Chang *et al.* [37] proved that hydrogen uptake and thereby formation of undesired hydrides can be caused by FIB preparation, where structural damage to the material supports higher hydrogen incorporation at the specimen surface [37]. Structural damage and thereby incorporation of hydrogen can be reduced by sample preparation at cryogenic temperature [37]. Therefore, cryogenic preparation was carried out in a xenon PFIB to compare the hydrogen uptake of the Mg-2Al-0.1Ca specimen to conventional preparation routes. Figure 3 shows the mass spectra of 3.5 million evaporated ions for specimens prepared conventionally in the gallium FIB and at cryogenic temperature in the xenon PFIB, evaporated in laser and voltage mode, respectively, in a range from 0 to 4 Da and 23 to 28 Da. Both spectra show peaks for hydrogen and its isotopes at 1, 2 and 3 Da. The appearance of peaks differs due to the employed atom probe microscopes. The cryogenically prepared specimen was measured in a straight flight path instrument and, thus, the detected mass spectrum is less confined (broader peaks) compared to the reflectron instrument, which was used for the conventionally prepared specimen. Nevertheless, the order of magnitude of hydrogen signal, i.e. the integrated peak area, are very similar for both preparation routes. In the range from 23 to 28 Da, only peaks for $Mg^+$ and $Mg_2^{2+}$ can be identified. Considering the isotope ratios fitting peaks from magnesium, the presence of significant amounts of MgH$_2^+$ (at 26 Da) can be ruled out [38].

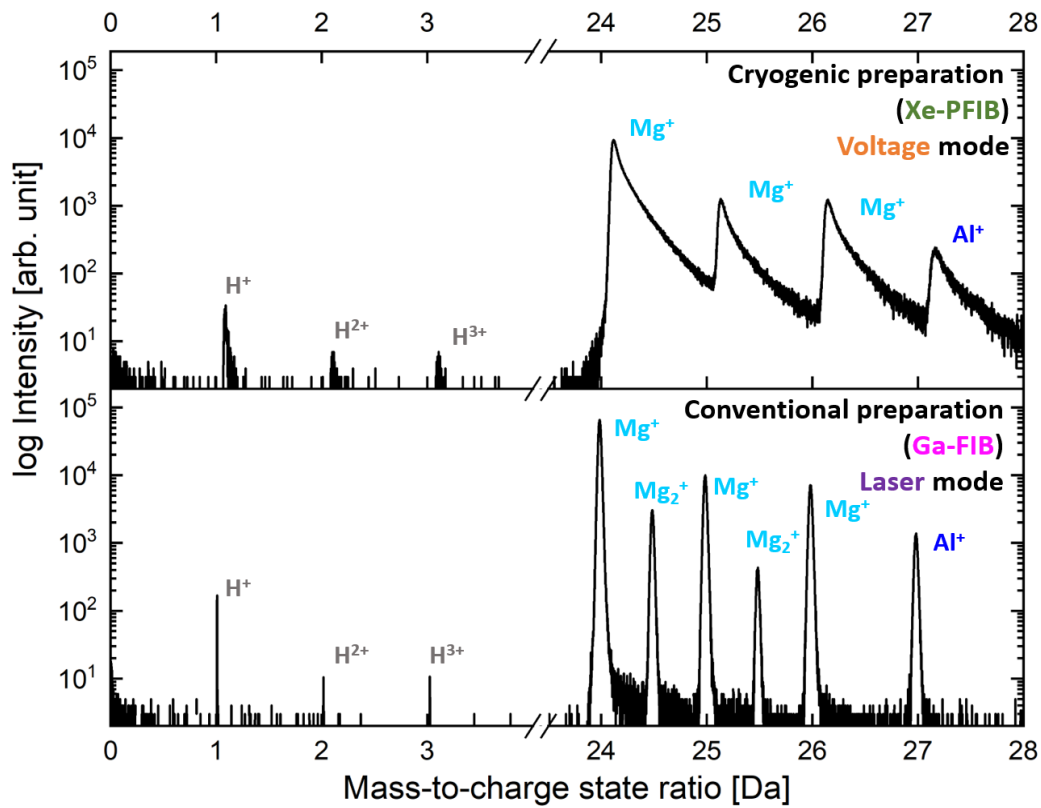

**Figure 3.** Mass spectra of Mg-2Al-0.1Ca after lift-out and sharpening with xenon PFIB and gallium FIB in a range from 0 to 4 Da with indication of hydrogen ions and from 23 to 28 Da with indication of Mg$^+$ and Mg$_2^{2+}$ and their isotopes, as well as Al$^+$. For both datasets 3.5 million ions were collected.

Based on these mass spectra the average hydrogen content for the initial 3.5 million evaporated ions were quantified. While the cryogenically prepared specimen contains around 1.5 at.%, the hydrogen content in the conventionally prepared sample is below 1 at.%. The composition profile from the cylindrical volume indicated in Figure 2 shows an increase in hydrogen to 1 at.% only for the first 100 to 200 nm of the reconstruction, decreasing below this value further into the bulk. Higher hydrogen contents in the apex region are expected since conventionally prepared specimens are exposed to air during transport from the FIB to the atom probe microscope. Moreover, the instrument-specific hydrogen content in the residual gas of the analysis chamber has an effect. Residual gas, containing hydrogen, can be adsorbed onto the specimen, leading to migration towards the apex region as well as eventual field desorption and ionization [39]. In addition, residual hydrogen in the atom probe chamber can produce additional $H_2O$, which can be absorbed at the specimen's surface [38]. Further recombination of hydrogen to $H_2$ and the formation of $OH^-$ and $Mg^{2+}$ ions can lead to reactions resulting in the detection of several combinations of $Mg_xH_y$ and $Mg_xO_yH_z$ molecules [38]. Here, preparation in a PFIB at cryogenic temperature did not show significant differences in incorporated hydrogen compared to the conventional preparation route.

3.2. Structure of the native oxide

Next, the crystal structure and orientation relationship of the interface between the native oxide and the Mg-2Al-0.1Ca bulk of a pristine atom probe specimen was characterized by aberration-corrected STEM. An ABF image is presented in Figure 4a and a HAADF image as well as EDX maps are shown in Figure 4b. It is evident that a thin oxide has formed on top of the Mg-2Al-0.1Ca alloy. The oxide scale is clearly observable in the STEM/EDX data, however, due to the geometry of the atom probe specimen and the projection effect [40], it is challenging to conclude whether aluminum is enriched in the oxide scale. Slight amounts of gallium impurities are still present and the calcium content was below the EDX detection limit.

The orientation relationships between bulk and native oxide were investigated and the (0002) basal plane of the bulk alloy is already oriented perpendicular to the atom probe specimen, Figure 4c and Figure 4d. The interface between (0002) lattice planes of the bulk and the native oxide at the apex of the atom probe specimen is presented in Figure 4e and it can be deduced from the images that 3-4 nm of MgO has formed with the orientation relationship MgO(111) // (Mg(0002) and Mg[2$\bar{2}$0] // Mg [1$\bar{1}$00], as schematically illustrated in Figure 4f. Similarly, at the side of the specimen a 3-4 nm native oxide is present with an orientation relationship MgO(106) // Mg(222) and MgO[200] // Mg[1$\bar{1}$00] (Figure 4g-h).

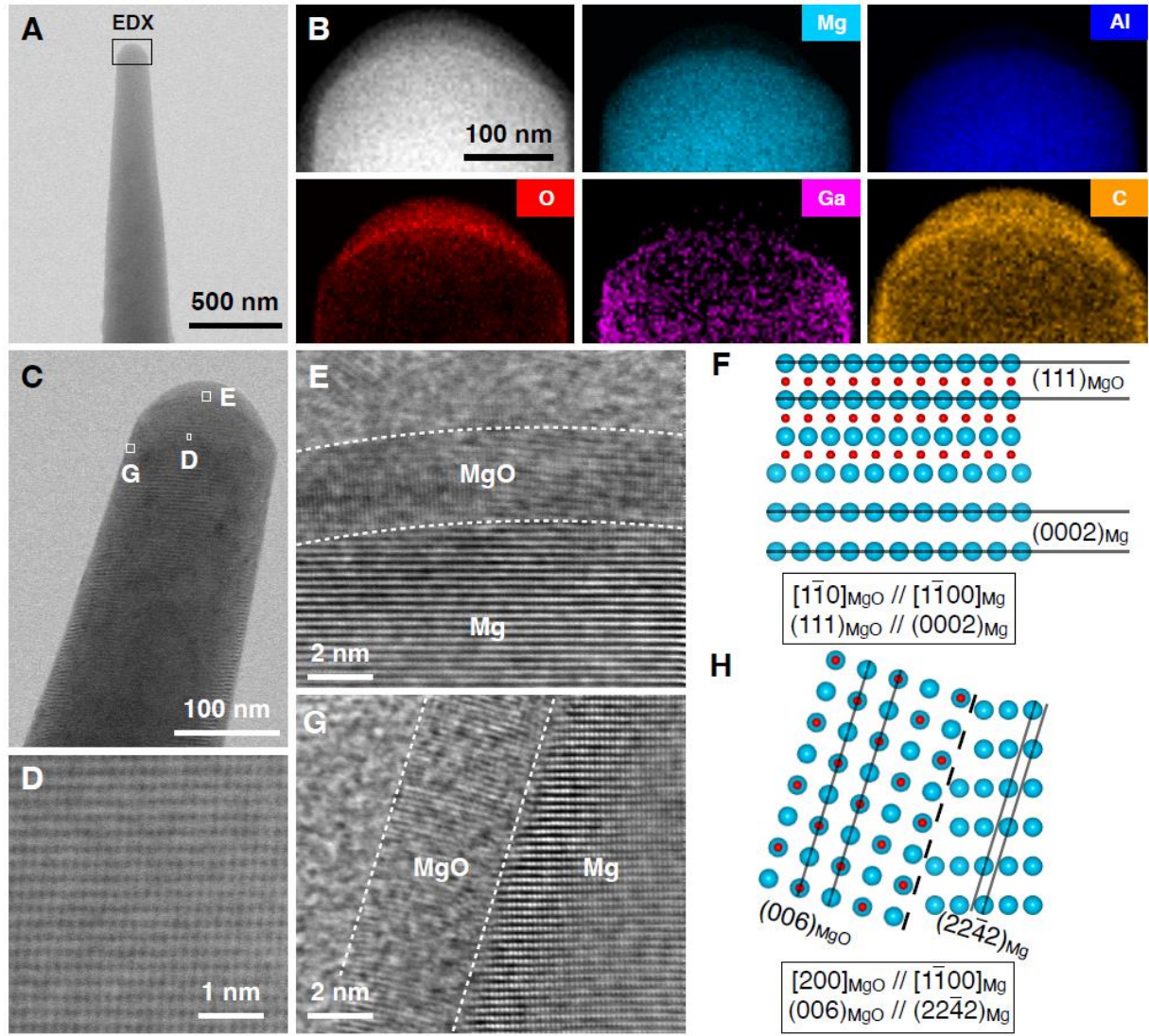

**Figure 4.** Aberration-corrected STEM investigation of the native oxide, which was formed on the pristine surface of the Mg-2Al-0.1Ca alloy. (a) ABF-STEM image of a cleaned atom probe specimen and (b) HAADF-STEM image as well as EDX maps from the region highlighted in (a). (c) Overview and (d) high-resolution ABF-STEM image of the specimen bulk. (e) High-resolution ABF-STEM image of the apex of the specimen and the orientation relationships between bulk and native oxide are illustrated in (f). (g) High-resolution ABF-STEM image of the side of the specimen and the orientation relationships between bulk and native oxide are illustrated in (h).

### 3.3. Chemical composition of the native oxide

Following the structural analysis with STEM, the chemical composition of the native oxide is characterized using APT. Figure 5 displays different mass spectra in a range from 0 to 75 Da for Mg-2Al-0.1Ca, oxidized for 12 h in ambient air and measured with laser and voltage mode, as well as a measurement of the magnesium reference after 11 days of oxidation in ambient air using voltage mode. All three spectra show the characteristic mass-to-charge state ratios for $Mg^+$ and $Mg^{2+}$ and their isotopes at 24, 25, 26 and 12, 12.5 and 13 Da, respectively. Further, for all three oxidized specimens, peaks of $MgO^+$ as well as $MgO^{2+}$ and their isotopes at around 40 and 20 Da can be seen. In case of Mg-2Al-0.1Ca, the mass spectra additionally display the peaks of $Al^+$ at 27 Da, as well as slight amounts of $Ga^+$ impurities at 69 Da. Significant differences in the peak shapes are caused by the flight paths of the atom probe microscopes. The laser mode measurements were done in an instrument with a reflectron, confining the time-of-flight of ion species and thereby improving the mass resolution [41]. Due to increased thermal input at the specimen, caused by the laser pulsing, many complex ions with overlapping peak positions form. Most prominent examples are $Mg_2^+$ and $Mg_2^{2+}$ molecular ions at 48 and 49, as well as 24.5 and 25.5 Da, respectively and several combinations of magnesium with oxygen and hydrogen, resulting in $Mg_xO_yH_z$ molecular ions.

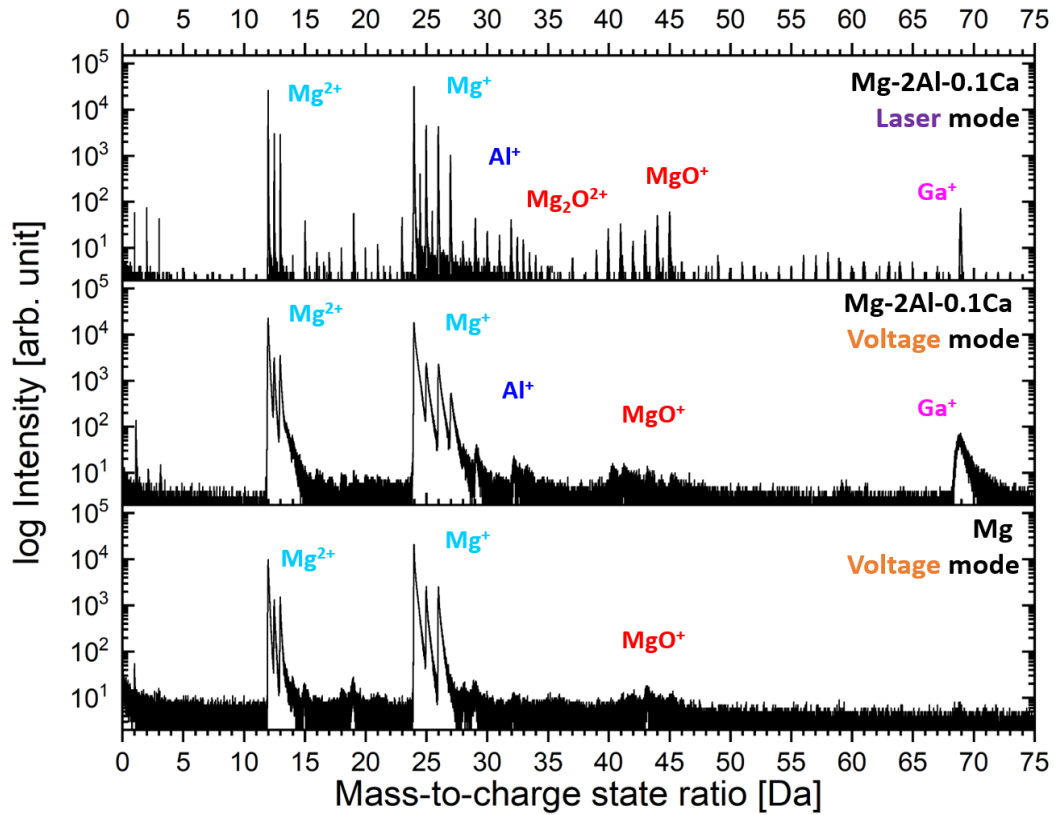

**Figure 5.** Mass spectra for Mg-2Al-0.1Ca oxidized in ambient air for 12 h measured in laser and voltage mode and the magnesium reference sample oxidized in air for 11 days measured in voltage mode. Peak positions for $Mg^{2+}$ and $Mg^+$ and their isotopes at 12 and 24 Da, respectively, $Al^+$ at 27 Da, $Mg_2O^{2+}$ and $MgO^+$ with their isotopes at 32 and 40 Da, respectively, as well as $Ga^+$ at 69 Da are indicated.

For the analysis of the native oxide, oxygen-containing molecular ions are of interest. While $O^+$ at 16 Da can be quantified, the majority of oxygen in the specimen is identified through the molecular ions in the form of $MgO_x$. Peak positions for $O_2^+$ and $Mg_2O^{2+}$ overlap, but can be resolved via the isotopic ratios and clearly identified as $Mg_2O^{2+}$. In contrast, for the voltage mode the only additional source of oxygen besides the $O^+$ peak at 16 Da is found in the $MgO^+$ peaks and their isotopes at around 40 Da. In laser mode, the additional thermal input of the laser causes the formation of different oxygen-containing $Mg_xO_yH_z$ molecular ions.

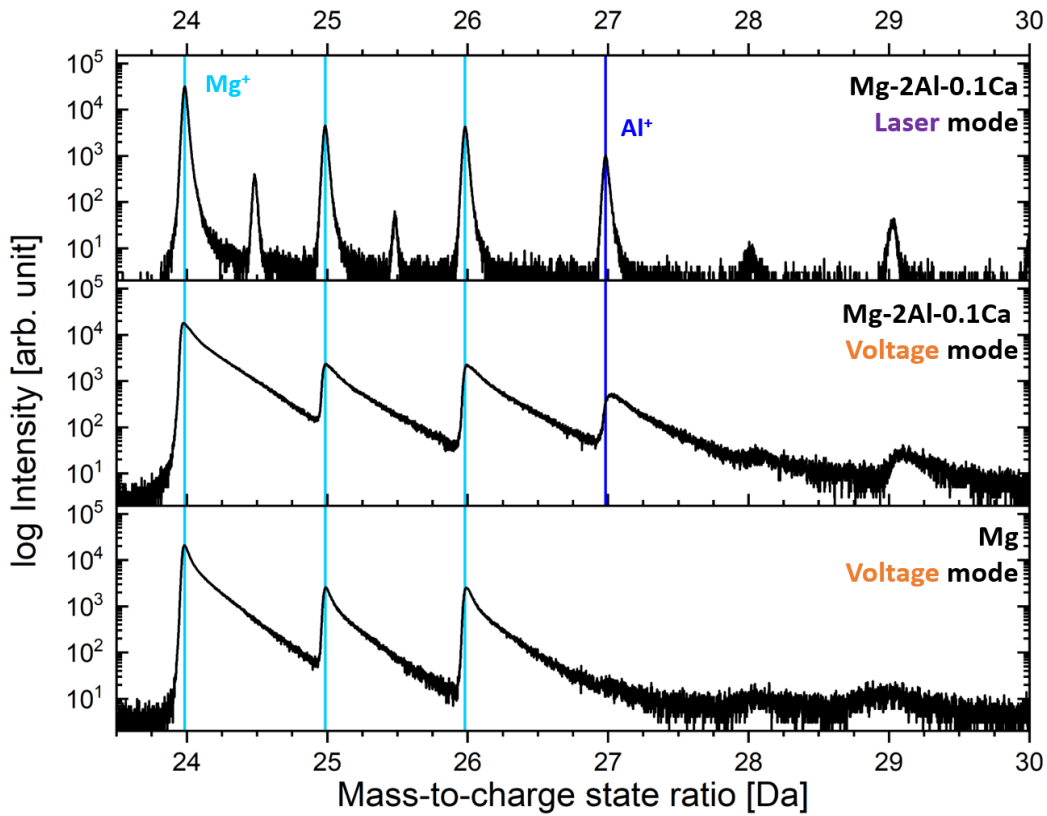

**Figure 6.** Detailed mass spectrum data in the range of 23 to 30 Da of Mg-2Al-0.1Ca oxidized in ambient air for 12 h measured in laser and voltage mode and magnesium reference sample oxidized in air for 11 days measured in voltage mode. Cyan and blue lines represent the abundance of $Mg^+$ and $Al^+$ isotopes, respectively.

The most crucial peaks to identify in the Mg-Al-Ca system are the ones for $Al^+$ and $Ca^+$ at 27 and 40 Da, respectively. The effect of aluminum alloying on the possible contribution to the native oxide scale is of particular interest for this study since previous works suggested the formation of an aluminum-rich zone at the interface between surface and bulk [21, 22]. For this reason, the peak at 27 Da, seen in Figure 6, is discussed next. While the peak at 27 Da is characteristic for $Al^+$ ions, recombined $MgH_x^+$-molecular ions can possess the same mass-to-charge state ratio. By comparison of the mass spectra from Mg-2Al-0.1Ca and the magnesium reference, it becomes evident, that the 27 Da peak only appears in samples containing aluminum and can therefore be identified as $Al^+$, consistent with STEM-EDX analysis (Figure 4b). Another remaining question is the presence of calcium. The distinction between $Ca^+$ and $MgO^+$, both with overlapping peaks at the position of 20 and 40 Da as well as their isotopes, is crucial to understand the possible influence of calcium on the native oxide formation. One indicator of whether $Ca^+$ or $MgO^+$ is detected is the intensity difference of ions with different charge states. $Ca^+$ (at 40 Da) and $Ca^{2+}$ (at 20 Da) have very similar evaporation fields of 18 and 19 V nm$^{-1}$, respectively, and therefore, the intensity of the peaks should be equally high. However, corresponding counts at 40 and 20 Da in Figure 5 are ~30 and ~10, respectively. Besides the main isotope, e.g. $Ca^+$ at 40 Da possesses a number of isotopes with very low abundances < 3%. While the measured isotope intensities do not match $Ca^+$ at 42, 43 and 44 Da, $MgO^+$ and $MgOH_y^+$ show much better agreement. Therefore, additional peaks at 41 and 43 Da can only be attributed to $MgOH^+$ and $MgOH_3^+$ molecular ions, respectively. Judging from the peaks present in the mass spectrum in Figure 5 neither of them can be confirmed as calcium. This finding is in good agreement with STEM-EDX data, where calcium was neither found in the bulk, nor in the native oxide. Based on density functional theory calculations of thermodynamic surface phase diagrams by Yang *et al.* [42], calcium incorporation induces large compressive strain to the Mg-Al-Ca system, which is released by segregation to the uppermost surface layer. Moreover, it is favorable to incorporate calcium at the surface since it lowers the surface energy of Mg(0001) [42]. However, since the specimens were taken from inside a grain, no grain boundaries are present, to which the calcium likely segregates in the as prepared samples. Hans *et al.* [43] correlated a bright field STEM image of an atom probe specimen from a Mg-Ca thin film with compositional APT data and calcium-rich regions were found at the grain boundaries. For these reasons, peaks in the range of 40 to 45 Da are attributed to $MgO^+$ and $MgOH_x^+$ molecular ions and their respective isotopes.

Side-by-side comparisons of the reconstructed APT specimens for oxidized Mg-2Al-0.1Ca, measured in laser and voltage mode, as well as oxidized magnesium, measured in voltage mode, are shown in Figure 7. It is evident that both laser and voltage mode can be used to capture the native oxide layer, which is clearly distinguishable from the bulk material. While side (Figure 7a-c) and top views (Figure 7d-f) give evidence for the presence of the native oxide by a visible enrichment of oxygen as well as hydrogen in the apex region, the oxide thickness of the reconstructions can be determined by the sliced 4 nm side views (Figure 7g-i). The outmost oxide layer thickness measured for the magnesium reference sample is ~1 nm (Figure 7i), while for Mg-2Al-0.1Ca measured in laser (Figure 7g) and voltage (Figure 7h) mode, a thickness of ~2 nm is obtained. Considering the effective spatial resolution of APT on the order of 1 nm as well as that during the onset of the atom probe measurement surface atoms are evaporated and not considered in the reconstruction, the measured native oxide scale thickness in APT of around 2 nm is in good agreement with the 3-4 nm thickness measured by aberration-corrected STEM. Aluminum-enrichment is clearly visible at the interface between bulk and native oxide of Mg-2Al-0.1Ca and aluminum clusters are also found in the bulk material despite the homogenization heat treatment during production of the alloy. The larger volume of aluminum and oxygen enriched regions and higher density of aluminum clusters in the specimen measured in voltage mode can be explained by the higher detection efficiency of 86% for the LEAP5000 compared to 36% for LEAP4000, leading to improved measurement statistics [44].

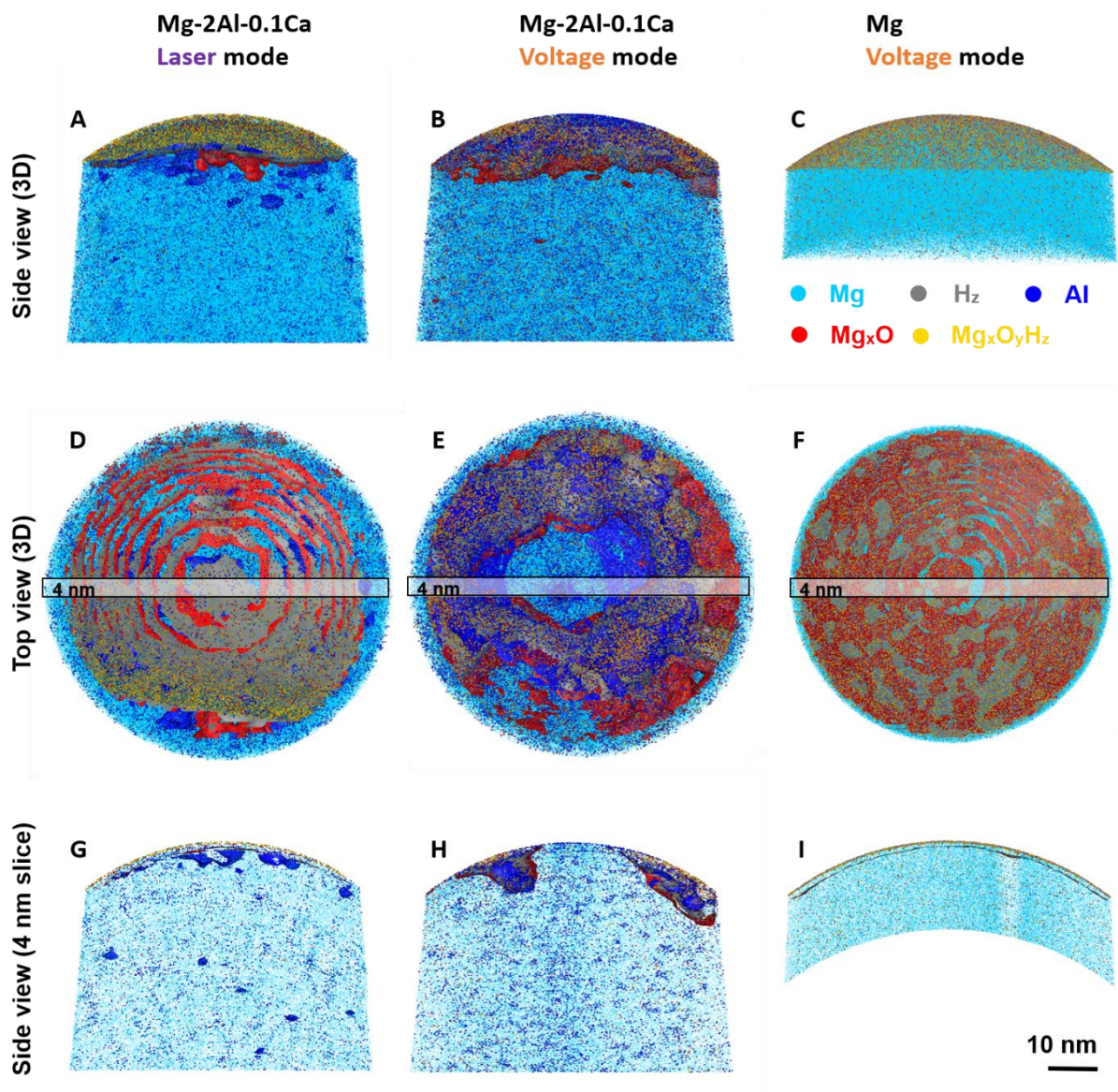

**Figure 7.** Side (a-c) and top view (d-f) of the full specimen from conventional preparation as well as side view on the 4 nm middle slice (g-i) for Mg-2Al-0.1Ca in laser and voltage mode and the magnesium reference in voltage mode with a color scheme for ions and molecules seen. Spherical depiction was used to indicate molecular ions and aluminum, while coherent shapes represent iso-concentration surfaces for the equivalent atoms, which were created with 9 at.% oxygen, 10 at.% hydrogen, and 6 at.% aluminum, respectively.

Besides visual analysis of atomic positions shown in Figure 7, a cylindrical concentration profile and a cumulative plot of number of ions vs. ion evaporation sequence, seen in Figure 8b and c respectively, provide evidence for aluminum at the interface between bulk and native oxide. The composition profile in Figure 8b shows a content of ~34 at.% oxygen in the surface region and ~66 at.% magnesium. The oxygen content strongly decreases at the interface and at a distance of 3 nm a maximum of ~20 at.% aluminum is obtained, while the amount of oxygen is only ~2 at.%. At a distance of 4 nm oxygen and aluminum levels are at ~1 and ~3 at.%, respectively, corresponding to the values for the bulk material. Since the cylinder profile describes the compositional evolution only locally, the global evolution of composition is addressed with the cumulative plot in Figure 8c. The number of detected Mg, Al and $Mg_2O$ ions is shown vs. the sequence of all detected ions and based on the different slopes three regions can be identified in agreement with the cylinder profile (Figure 8b). Up to ion sequence 1200 there is a steep increase of detected $Mg_2O$ ions, which can be correlated to the native oxide at the surface. In between the ion sequence of 900 to 2100 an increase of detected Al ions is observed, while the slope of $Mg_2O$ stagnates. Hence, this region can be attributed to an aluminum-rich interface. From ion sequence 2100, both, the number of detected $Mg_2O$ and Al ions remains comparatively constant, indicating the transition from the aluminum-rich interface to bulk material.

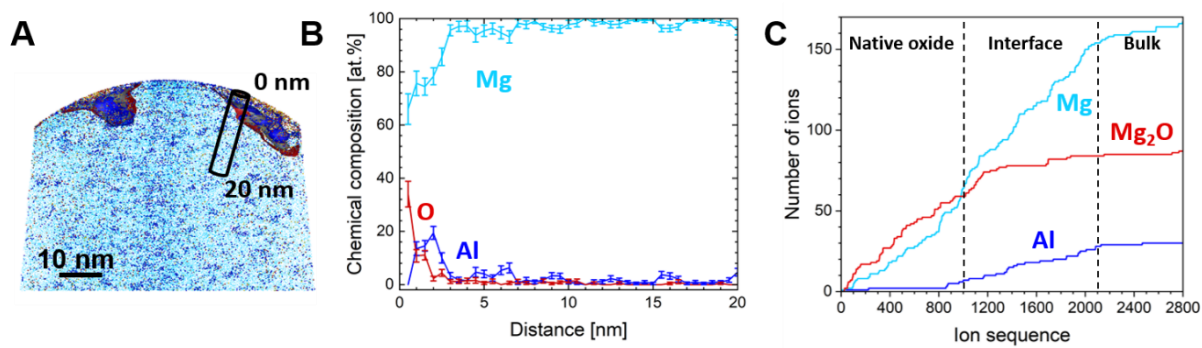

**Figure 8.** (a) 4 nm slice of Mg-2Al-0.1Ca evaporated in voltage mode with indication of the 4x4x20 nm cylindrical profile. (b) Concentration profile of cylindrical region indicated in (a) for magnesium, aluminum and oxygen. (c) Cumulative plot of number of ions vs. ion sequence for Mg, $Mg_2O$ and Al ions.

Based on the Al enrichment at the interface between bulk and native oxide, an inward growth of the oxide scale can be derived. The presence of aluminum clusters in the bulk material supports this notion. Due to the inward growth of the native oxide, aluminum as well as aluminum clusters are pushed deeper into the bulk, forming the enriched region at the interface between bulk and native oxide. Since magnesium oxide forms a protective scale, meaning the oxidation rate decreases with time, it can be assumed that the native oxide of Mg-2Al-0.1Ca is stable upon longer exposure times [45, 46]. Thus, it can be conceived that the aluminum-rich interface is saturated, when the native oxide stabilizes. The here identified aluminum-rich interface provides a baseline for future studies on the corrosion behavior of aluminum-containing magnesium-based alloys.

While the native oxide could be clearly identified by local chemical composition analysis, the measured oxygen content through APT is limited by the measurement accuracy. The highest measured O content in the native oxide is ~34 at.%. However, STEM-EDX measurements and high-resolution STEM images identified the present oxide layer as MgO, implying that an O content of ~50 at.% would be expected. The ion yield in MgO at the highest evaporation field was found to be not stoichiometric and explained by differences in the evaporation fields for magnesium and oxygen [47]. For non-stoichiometric ion yields, different scenarios of oxygen loss during the measurement are discussed by Karahka *et al.* [47] for MgO and ZnO oxides. First, the migration of negative $O^-$ ions down the surface of the specimen results in eventual thermal desorption and the formation of $O_2$. From the oxygen accumulation either molecular oxygen-containing ions can form at the surface or $O_2$ gas forms due to associative desorption [47]. The formation of non-detectable neutral $O_2$ molecules reduces the detection efficiency of oxygen [48].

Finally, the presence of a hydroxylated layer on top of the MgO native oxide is discussed based on the atom probe data shown in Figure 7. As mentioned in section 3.1, specimen preparation did not have a significant influence on the hydrogen content in magnesium. In case of the native oxide $Mg_xO_yH_z$ molecular ions are detected with up to ~ 50 at.% hydrogen in the Mg-2Al-0.1Ca specimen, evaporated in laser mode (Figure 7a), as well as in magnesium, evaporated in voltage mode (Figure 7c). These measured hydrogen contents can be attributed to different factors. It has been shown for magnesium that the exposure of the formed native oxide to water vapor in ambient air causes hydroxylation to $Mg(OH)_2$ by chemisorption [49]. As the here investigated magnesium specimen was exposed to ambient air for 11 days, it can be expected the longer exposure time leads to a higher hydroxylation rate, that can be quantified by the higher amount of $Mg_xO_yH_z$ molecular ions. As discussed above, the Mg-2Al-0.1Ca specimen, evaporated in laser mode, is prone to form molecular ions with residual gas in the chamber due to thermal energy input caused by the laser pulsing. While the Mg-2Al-0.1Ca specimen, measured in voltage mode, was only exposed to air for 12 hours, the hydrogen uptake is not as high as in the Mg reference sample. Hence, hydroxylation cannot be inferred from the here presented data, but the stacking sequence of a $Mg(OH)_2$ hydroxylated layer on top of the MgO native oxide, followed by an aluminum-rich interface and finally the bulk material appears to be in good agreement with a study by Felten *et al.* [49], which also indicated the protective nature of the native oxide scale. It has been further reported by Aliramaji *et al.* [46] that the MgO scale is a natural passivation layer, forming within the first minutes of atmosphere exposure and inhibiting the further oxygen incorporation upon air exposure.

## 4. Conclusions

The native oxide, formed on the basal (0001) plane of Mg-2Al-0.1Ca has been studied regarding its structure and chemical composition, combining STEM and APT data. To evaluate the incorporation of hydrogen during specimen preparation, gallium FIB at ambient temperature and xenon PFIB at cryogenic temperature with a vacuum transfer from the PFIB to the atom probe system were employed. The comparison of mass spectra showed no significant reduction in hydrogen incorporation for the cryogenic preparation, but rather that the measured hydrogen content in Mg seems to be governed by instrument-specific levels of residual gas.

STEM investigations revealed the formation of a 3-4 nm thick layer MgO. At the apex of the atom probe specimen where Mg is terminated on the basal plane, the orientation relationship MgO(111) // Mg(0002) and MgO[$2\bar{2}0$] // Mg [$1\bar{1}00$] is observed. Moreover, at the sides of the specimen with the termination of Mg($22\bar{4}2$), another orientation relationship MgO (006) // Mg($22\bar{4}2$) and MgO[200] // Mg[$1\bar{1}00$] is observed. An aluminum-rich region forms between bulk and native oxide upon exposure to ambient air as revealed by both laser-assisted and voltage field evaporation. The aluminum enrichment of up to ~20 at.% at the interface between bulk and native oxide is consistent with the inward growth of the oxide scale. Calcium has not been identified due to very low content and possibly overlapping peaks in the mass spectrum.

It is demonstrated that the approach of combining STEM and APT enables unravelling the structure and composition of the native oxide scale, formed on the surface of magnesium-based alloys. Hence, these findings provide the baseline for studying complex corrosion phenomena of magnesium-based alloys during high temperature oxidation and immersion testing in the future.

**Acknowledgments**

This work was supported by Deutsche Forschungsgemeinschaft (DFG) within the Collaborative Research Center SFB 1394 „Structural and Chemical Atomic Complexity – From Defect Phase Diagrams to Materials Properties" (project ID 409476157). Ingrid McCarroll is grateful for funding from the ERC – SHINE (771602) and DFG through the award of the Leibniz Prize 2020. Siyuan Zhang acknowledges funding from the DFG under the framework of SPP 2370 (Project number: 502202153). Hauke Springer acknowledges financial support from the Heisenberg-Programm of the DFG (project number 416498847). Uwe Tezins, Andreas Sturm, Tim Schwarz, and Christian Broß are acknowledged for their support of the FIB, cryo-FIB, and APT facilities at Max-Planck-Insitut für Eisenforschung GmbH (MPIE). Furthermore, the authors thank Dierk Raabe, Baptiste Gault and Narasimha Sasidhar Kasturi from MPIE for fruitful discussions.

**Conflict of Interest**

The authors declare no conflict of interest.

**Data Availability Statement**

The data that support the findings of this study are available from the corresponding author upon reasonable request.

**References**

[1] J. Hirsch, T. Al-Samman, *Acta Mater.* **2013**, *61*, 818-843.
https://doi.org/10.1016/j.actamat.2012.10.044

[2] H. Friedrich, S. Schumann, *J. Mater. Process. Technol.* **2001**, *117*, 276-281.
10.1016/S0924-0136(01)00780-4

[3] B. R. Powell, A. A. Luo, P. E. Krajewski, in *Advanced Materials in Automotive Engineering* (Ed: J. Rowe), Woodhead Publishing Limited, Cambridge, UK **2012**, Ch. 7, pp. 150-209.
10.1533/9780857095466.150

[4] B. R. Powell, P. E. Krajewski, A. A. Luo, in *Materials, Design and Manufacturing for Lightweigth Vehicles* (Ed: P. K. Mallick), Woodhead Publishing Limited, Cambridge, UK **2010**, Ch. 4, pp. 114-173.
10.1533/9781845697822.1.114

[5] A. A. Luo, *JOM* **2002**, *54*, 42-48.
10.1007/BF02701073

[6] D. Kumar, R. K. Phanden, L. Thakur, *Mater. Today: Proc.* **2021**, *38*, 359-364.
10.1016/j.matpr.2020.07.424

[7] H. Hu, A. Yu, N. Li, J. E. Allison, *Mater. Manuf. Process.* **2003**, *18*, 687-717.
10.1081/AMP-120024970

[8] F. Czerwinski, *JOM* **2012**, *64*, 1477-1483.
10.1007/s11837-012-0477-z

[9] Q. Tan, A. Atrens, N. Mo, M.-X. Zhang, *Corros. Sci.* **2016**, *112*, 734-759.
10.1016/j.corsci.2016.06.018

[10] V. Fournier, P. Marcus, I. Olefjord, *Surf. Interface Anal.* **2002**, *34*, 494-497.
10.1002/sia.1346

[11] F. Czerwinski, *Acta Mater.* **2002**, *50*, 2639-2654.
10.1016/s1359-6454(02)00094-0

[12] G. Song, A. Atrens, *Adv. Eng. Mater.* **1999**, *1*, 11-33.
10.1002/(SICI)1527-2648(199909)1:1<11::AID-ADEM11>3.0.CO;2-N

[13] M. Liu, D. Qiu, M.-C. Zhao, G. Song, A. Atrens, *Scripta Mater.* **2008**, *58*, 421-424.
10.1016/j.scriptamat.2007.10.027

[14] Q. Tan, N. Mo, B. Jiang, F. Pan, A. Atrens, *Corros. Sci.* **2017**, *122*, 1-11.
https://doi.org/10.1016/j.corsci.2017.03.023

[15] A. Drynda, T. Hassel, R. Hoehn, A. Perz, F. W. Bach, M. Peuster, *J. Biomed. Mater. Res.* **2010**, *93*, 763-775.
10.1002/jbm.a.32582

[16] B.-S. You, W.-W. Park, I.-S. Chung, *Scripta Mater.* **2000**, *42*, 1089-1094.
10.1016/S1359-6462(00)00344-4

[17] S. Mathieu, C. Rapin, J. Steinmetz, P. Steinmetz, *Corros. Sci.* **2003**, *45*, 2741-2755.
10.1016/s0010-938x(03)00109-4

[18] S. Feliu Jr., J. C. Galván, A. Pardo, M. C. Merino, R. Arrabal, *The Open Corrosion Journal* **2010**, *3*, 80-91.
10.2174/1876503301003010080

[19] B.-S. You, W.-W. Park, I.-S. Chung, *Mater. Trans.* **2001**, *42*, 1139-1141.
10.2320/matertrans.42.1139

[20] S.-H. Ha, Y.-O. Yoon, B.-H. Kim, H.-K. Lim, T.-W. Lee, S.-H. Lim, S. K. Kim, *Int. J. Met.* **2019**, *13*, 121-129.
10.1007/s40962-018-0234-3

[21] F. Hermann, F. Sommer, H. Jones, R. G. J. Edyvean, *J. Mater. Sci.* **1989**, *24*, 2369-2379.
10.1007/bf01174498

[22] L. P. H. Jeurgens, M. S. Vinodh, E. J. Mittemeijer, *Acta Mater.* **2008**, *56*, 4621-4634.
10.1016/j.actamat.2008.05.020

[23] Y. Amouyal, G. Schmitz, *MRS Bull.* **2016**, *41*, 13-18.
10.1557/mrs.2015.313

[24] D. Blavette, A. Bostel, J. M. Sarrau, B. Deconihout, A. Menand, *Nature* **1993**, *363*, 432-435.
10.1038/363432a0

[25] R. Haydock, D. R. Kingham, *Phys. Rev. Lett.* **1980**, *44*, 1520-1523.
10.1103/PhysRevLett.44.1520

[26] E. W. Müller, *Phys. Rev.* **1956**, *102*, 618-624.

10.1103/PhysRev.102.618

[27] M. K. Miller, R. G. Forbes, *Atom-Probe Tomography - The Local Electrode Atom Probe*, Springer, New York, NY, USA **2014**.

[28] B. Gault, M. P. Moody, J. M. Cairney, S. P. Ringer, Atom Probe Microscopy, Springer, New York, NY, USA **2012**.

[29] F. De Geuser, B. Gault, *Acta Mater.* **2020**, *188*, 406-415.
10.1016/j.actamat.2020.02.023

[30] F. De Geuser, B. Gault, A. Bostel, F. Vurpillot, *Surf. Sci.* **2007**, *601*, 536-543.
10.1016/j.susc.2006.10.019

[31] K.N. Sasidhar, H. Khanchandani, S. Zhang, A. Kwiatkowski da Silva, C. Scheu, B. Gault, D. Ponge, D. Raabe, *Corros. Sci.* **2023**, *211*, 110848.
10.1016/j.corsci.2022.110848

[32] F. Moens, I. C. Schramm, S. Konstantinidis, D. Depla, *Thin Solid Films* **2019**, *689*, 137501.
10.1016/j.tsf.2019.137501

[33] S. Zhang, Z. Xie, P. Keuter, S. Ahmad, L. Abdellaoui, X. Zhou, N. Cautaerts, B. Breitbach, S. Aliramaji, S. Korte-Kerzel, M. Hans, J.M. Schneider, C. Scheu, *Nanoscale* **2022**, *14*, 18192-18199.
10.1039/D2NR05505H

[34] K. Thompson, D. Lawrence, D. J. Larson, J. D. Olson, T. F. Kelly, B. Gorman, *Ultramicroscopy* **2007**, *107*, 131-139.
10.1016/j.ultramic.2006.06.008

[35] M. Herbig, P. Choi, D. Raabe, *Ultramicroscopy* **2015**, *153*, 32-39.
10.1016/j.ultramic.2015.02.003

[36] S. Zhang, C. Scheu, *Microscopy* **2018**, *67*, i133-i141.
10.1093/jmicro/dfx091

[37] Y. Chang, W. Lu, J. Guenolé, L. T. Stephenson, A. Szczepaniak, P. Kontis, A. K. Ackerman, F. F. Dear, I. Mouton, X. Zhong, S. Zhang, D. Dye, C. H. Liebscher, D. Ponge, S. Korte-Kerzel, D. Raabe, B. Gault, *Nat. Commun.* **2019**, *10*, 942.
10.1038/s41467-019-08752-7

[38] I. E. McCarroll, D. Haley, S. Thomas, M. S. Meier, P. A. J. Bagot, M. P. Moody, N. Birbilis, J. M. Cairney, *Corros. Sci.* **2020**, *165*, 108391.
10.1016/j.corsci.2019.108391

[39] I. Mouton, A. J. Breen, S. Wang, Y. Chang, A. Szczepaniak, P. Kontis, L. T. Stephenson, D. Raabe, M. Herbig, T. Ben Britton, B. Gault, *Microsc. Microanal.* **2019**, *25*, 481-488.
10.1017/S143192761801615X

[40] F. De Geuser, B. Gault., *Microsc. Microanal.* **2017**, 23(2), 238-246.
10.1017/S1431927616012721

[41] P. Clifton, T. Gribb, S. Gerstl, R. M. Ulfig, D. J. Larson, *Microsc. Microanal.* **2008**, *14*, 454-455.
10.1017/S1431927608087217

[42] J. Yang, K. B. Sravan Kumar, M. Todorova, J. Neugebauer, *(Preprint)* arxiv:2207.09809, v2, submitted: February, **2023**.

[43] M. Hans, P. Keuter, A. Saksena, J. A. Sälker, M. Momma, H. Springer, J. Nowak, D. Zander, D. Primetzhofer, J. M. Schneider, *Sci. Rep.* **2021**, *11*, 17454.
10.1038/s41598-021-97036-6

[44] D. Beinke, C. Oberdorfer, G. Schmitz, *Ultramicroscopy* **2016**, *165*, 34-41.
10.1016/j.ultramic.2016.03.008

[45] M. Kurth, P. C. J. Graat, E. J. Mittemeijer, *Thin Solid Films* **2006**, *500*, 61-69.
10.1016/j.tsf.2005.11.044

[46] S. Aliramaji, P. Keuter, D. Neuß, M. Hans, D. Primetzhofer, D. Depla, J. M. Schneider, *Materials* **2023**, *16*, 414.
10.3390/ma16010414

[47] M. Karakha, Y. Xia, J. J. Kreuzer, *Appl. Phys. Lett.* **2015**, *107*, 062105.
10.1063/1.4928625

[48] B. Gault, D. W. Saxey, M. W. Ashton, S. B. Sinnott, A. N. Chiaramonti, M. P. Moody, D. K. Schreiber, *New J. Phys.* **2016**, *18*, 033031.
10.1088/1367-2630/18/3/033031

[49] M. Felten, J. Nowak, O. Beyss, P. Grünewald, C. Motz, D. Zander, *Corros. Sci.* **2023**, *212*, 110925.
10.1016/j.corsci.2022.110925